\documentclass[twocolumn,showpacs,preprintnumbers,amsmath,amssymb]{revtex4}
\usepackage{graphicx}% Include figure files
\usepackage{dcolumn}% Align table columns on decimal point
\usepackage{bm}% bold math

\begin{document}

%\preprint{APS/123-QED}

\title{Spin-Charge Soldering from Tensor Higgs mechanism}% Force line breaks with \\

\author{M. Cristina Diamantini}
%\altaffiliation[Also at ]{Theory Division, CERN, CH-1211 Geneva 23, Switzerland}%Lines break automatically or can be forced with \\
\email{cristina.diamantini@pg.infn.it}
\affiliation{%
NiPS Laboratory, INFN and Dipartimento di Fisica, University of Perugia, via A. Pascoli, I-06100 Perugia, Italy
}%

%\author{Pasquale Sodano}
%\altaffiliation[Also at ]{Perimeter Institute of Theoretical Physics 31, Caroline St. North, Waterloo, Ontario N2L2Y5, Canada}%Lines break automatically or can be forced with \\
%\author{Second Author}%
%\email{pasquale.sodano@pg.infn.it}
%\affiliation{%
%INFN and Dipartimento di Fisica, University of Perugia, via A. Pascoli, I-06100 Perugia, Italy
%}%

\author{Carlo A. Trugenberger}
%\altaffiliation[Also at ]{Theory Division, CERN, CH-1211 Geneva 23, Switzerland}%Lines break automatically or can be forced with \\
%\author{Second Author}%
\email{ca.trugenberger@bluewin.ch}
\affiliation{%
SwissScientific, chemin Diodati 10, CH-1223 Cologny, Switzerland
}%

%\author{M. Cristina Diamantini}
%\homepage{http://www.Second.institution.edu/~Charlie.Author}
%\affiliation{
%Second institution and/or address\\
%This line break forced% with \\
%}%

\date{\today}% It is always \today, today,
             %  but any date may be explicitly specified

\begin{abstract}
Spin-charge separation, a crucial ingredient in 2D models of strongly correlated systems, in mostly considered in condensed matter applications. In this paper we present a relativistic field-theoretic model in which charged particles of spin 1/2 emerge by soldering
spinless charges and magnetic vortices in a confinement quantum phase transition modelled as a tensor Higgs mechanism. The model involves two gauge fields, a  vector one and a two-form gauge field interacting by the topological BF term. When this tensor gauge symmetry is spontaneously broken charges are soldered to the ends of magnetic vortices and thus confined by a linear potential. If the vector potential has a topological $\theta $-term with value $\theta = \pi$, the constituents of this "meson" acquire spin 1/2 in this transition. 
\end{abstract}
\pacs{11.10.-z,11.15.Wx,73.43.Nq}
\maketitle

%\section{Introduction}
The concept of spin-charge separation is one of the guiding principles of the modern approach to strongly correlated, low-dimensional systems \cite{ave}. The idea is that, in specific ground states, the electron is fractionalized into two "constituent" quasi-particles, the chargeon (holon), carrying only the charge degree of freedom and the spinon, carrying only the spin degree of freedom. The two quasi-particles interact via emergent gauge fields: the electron is reconstituted when the gauge interaction becomes strong enough to cause confinement. 

This mechanism was shown by Haldane \cite{hal} to be a generic feature of 1D metallic systems. Moreover, the idea of electron fractionalization is thought to play a crucial role in the physics of the high-$T_c$ cuprates. Indeed, spin-charge separation seems to be an unavoidable characteristics of the 2D $tJ$ model \cite{weng, ye} of the doped Mott insulators, capturing the essential physics of high-$T_c$ superconductivity \cite{wensach}. Electron fractionalization in these models leads to new quantum orders not characterized by symmetry breaking \cite{sen1}. 

A key ingredient of the spin-charge separation idea in 2D is the representation of chargeons and spinons as a two-fluid model with mutual Chern-Simons interactions \cite{wensach}, a picture that can be analytically derived from the $tJ$ model \cite{weng, ye}. Mixed Chern-Simons fluids as representations of condensed matter systems where first introduced in \cite{dst1}, where it was shown that they capture all the essential physics of 2D Josephson junction arrays. In the same paper it was also pointed out that this two-fluid construction can be generalized to 3D, the topological interaction being encoded in what is known as the $BF$ term \cite{birmi}. Fractionalization in higher dimensions was further studied in \cite{sent} in the context of microscopic  two-fluid lattice models.

Spin-charge separation has been studied mainly in the framework of condensed matter models. Much less is known about the possibility of such fractionalisation in the context of relativistic field theory. In this paper we present a simple model in which we show that charged particles with spin 1/2 emerge in a model of spinless charges and magnetic vortices in a confinement quantum phase transition corresponding to a tensor version of the Higgs mechanism, in which an antisymmetric tensor field "eats up" the electromagnetic vector gauge field \cite{que}. As a consequence of this mechanism, charges are soldered to the endpoints of the vortex strings and become thus confined by a linear potential. In presence of a topological $\theta $-term $\theta = \pi$ in the electromagnetic action, the constituents of this "meson" acquire spin 1/2. The spin arises in a subtle way: the magnetic vortex is characterised by a conserved antisymmetric tensor, its (pseudo-) vector charges carry thus 2 degree of freedom that can be taken to describe a spin 1/2 representation of SU(2). In the tensor broken phase, particle-antiparticle pairs are connected by a magnetic vortex 
with a topological term representing the self-intersection number of its world-surface and the spin 1/2 representation is effectively attached to the particles at the end of this string \cite{paw}.  We stress that we study here only the kinematics of this quantum phase transition, we cannot present an explicit model of confinement via spontaneous tensor gauge symmetry breaking at this point. 

Let us consider point charges and closed vortices described by vector and antisymmetric pseudo-tensor currents (we shall consider Euclidean 4D space with the conventions $c=1$ and $\hbar=1$)
\begin{eqnarray}
J_{\mu} &&= \int_C d\tau \ {dx_{\mu}(\tau)\over d\tau} \ \delta^4 \left( {\bf x} - {\bf x(\tau)} \right) \ ,
\nonumber \\
\Phi_{\mu \nu} &&= {1\over 2} \int _S d^2 \sigma \ X_{\mu \nu} ({\sigma}) \ \delta^4 \left( {\bf x} - {\bf x(\sigma)} \right) 
\nonumber \\
X_{\mu \nu} &&= \epsilon^{ab} {\partial x_{\mu} \over \partial \sigma^a} {\partial x_{\nu} \over \partial \sigma^b} \ ,
\label{one}
\end{eqnarray}
where $C$ and $S$ are closed curves and compact surfaces parametrized by ${\bf x(\tau)}$ and ${\bf x(\sigma)}$ respectively. 
The charge current has the standard coupling $ iA_{\mu} J_{\mu}$ to the electromagnetic gauge vector $A_{\mu}$.  As usual, vortices couple to an antisymmetric gauge field $b_{\mu \nu}$ as $ib_{\mu \nu} \Phi_{\mu \nu}$. 

The two currents are conserved, $\partial_{\mu }J_{\mu} =0$, $\partial_{\mu } \Phi_{\mu \nu} = \partial_{\nu } \Phi_{\mu \nu}=0$. As a consequence, the two gauge fields are invariant under the usual vector gauge transformations
\begin{equation}
A_{\mu} \to A_{\mu} + \partial _{\mu} \lambda \ ,
\label{two}
\end{equation}
and under the tensor gauge transformations
\begin{equation}
b_{\mu \nu} \to b_{\mu \nu} + \partial_{\mu} \eta_{\nu} - \partial_{\nu} \eta_{\mu} \ .
\label{three}
\end{equation}
Note that , due to $\partial_i\Phi_{0i} =0$ the vortex vector charges carry two degrees of freedom, which are thus ideally suited to describe a spin 1/2 representation of SU(2). As we now show, this spin 1/2 representation will eventually be attached to the charges at each end of an open vortex in the spontaneously broken tensor symmetry phase. 

Let us now consider the most general (relativistic) gauge field action compatible with the two separate gauge invariances (\ref{two}) and (\ref{three}). Keeping only relevant and marginal terms, this is given by 
\begin{eqnarray}
S &&= {ik \over 16 \pi} \int d^4x \ b_{\mu \nu} \epsilon_{\mu \nu \alpha \beta} F_{\alpha \beta} +
{i\theta \over 16 \pi^2} \int d^4x \ F_{\mu \nu} \tilde F_{\mu \nu} +
\nonumber \\
&&+  {1\over 4 e^2 } \int d^4x \ F_{\mu \nu} F_{\mu \nu}  \ ,
\label{four}
\end{eqnarray}
where $F_{\mu \nu} = \partial_{\mu}A_{\nu}-\partial_{\nu}A_{\mu}$ is the field strength associated with $A_{\mu}$, $\tilde F_{\mu \nu} = (1/2) \epsilon_{\mu \nu \alpha \beta} F_{\alpha \beta}$ its dual, $e$ the charge unit and $k$ and $\theta$ dimensionless parameters. 

The first two terms in this action are purely topological terms: the first is called generically the $BF$ term \cite{birmi} and represents a generalization to 3D of the mutual Chern-Simons term in 2D. It preserves the $P$ and $T$ symmetries if the two-form gauge field is a pseudo-tensor, as it should be when coupling to pseudo-tensor vortices. The second is the famed $\theta$-term of axion electrodynamics \cite{wil}. The parameter $\theta $ is an angle variable with periodicity $2\pi$, the partition function being invariant under the shift $\theta \to \theta +2\pi$. The $\theta$-term breaks generically the $P$ and $T$ symmetries: these are however restored when $\theta$ is quantized: $\theta = n \pi$, $n \in \mathbb{ Z}$. Thus there are only two possible $\theta $ values compatible with the $P$ and $T$ symmetries: $\theta = 0$ and $\theta = \pi$. The third term in the action (\ref{four}) is the standard Maxwell action for the vector gauge field $A_{\mu}$. 

In (\ref{four}) the mixed components $b_{0i}=-b_{i0}$ play the role of non-dynamical Lagrange multipliers, leaving three dynamical degrees of freedom. The gauge invariance ({\ref{three}), however eliminates two of these, since there are two independent gauge parameters $\eta_i$ (the other one being eliminated by the equivalence $\eta_i\equiv \eta_i+ \partial _i \rho $), leaving thus one overall degree of freedom. The gauge invariant antisymmetric tensor $b_{\mu \nu}$ describes thus a pseudo-scalar degree of freedom.

Exactly as its Chern- Simons counterpart in 2D \cite{jackiw}, the $BF$ term gives rise to a gauge invariant mass for both the vector field $A_{\mu }$ and the pseudo-scalar described by $b_{\mu \nu}$ \cite{bow}.  This can be seen easily by adding to the action ,as a regulator, the gauge invariant, but infrared irrelevant kinetic term for the antisymmetric pseudo-tensor
\begin{equation}
S\to S+S_{\rm reg}, \ S_{\rm reg} = {1\over 12 \Lambda^2} \int d^4 x \ h_{\mu \nu \alpha}h_{\mu \nu \alpha} \  ,
\label{five}
\end{equation}
where $h_{\mu \nu \alpha} =  \partial_\mu b_{\nu \alpha} + \partial_\nu b_{ \alpha \mu} +\partial_\alpha b_{\mu \nu }$ is the field strength associated with the two-form gauge field $b_{\mu \nu}$ and $\Lambda$ is a mass parameter of the order of the ultraviolet cutoff $\Lambda_0$. This regulator term makes all quadratic kernels well defined by inducing a mass $m=e\Lambda k/4\pi $ for all fields. This is the anticipated topological, gauge invariant mass \cite{bow} that is the 3D analogue of the famed Chern-Simons topological mass \cite{jackiw}. At energies well below this scale, the equation of motion for the antisymmetric pseudo-tensor reduces to $\Phi_{\mu \nu} = (k/8\pi) \tilde F_{\mu \nu}$, which shows that the vortices become magnetic via the BF term: $\Phi_{0i} = (k/4\pi) B_i$. The action (\ref{four}) coupled to (\ref{one}) describes thus screened charges and magnetic vortices interacting only via Aharonov-Bohm phases. When $\theta = \pi$ is chosen, this is exactly the physics of strong topological insulators \cite{topins}, the surface metallic modes being described by the coupling 
\begin{equation}
\int_{\partial } d^3x {k\over 8\pi } A_{\mu} n_{\nu} \epsilon_{\mu \nu \alpha \beta} \partial_{\alpha} \chi_{\beta} \ ,
\label{six}
\end{equation} 
with the additional modes $\chi_{\mu}$ necessary to restore full tensor gauge invariance of the BF term on the boundary of an open hyper surface with normal vector ${\bf n}$, exactly as in the case of the well known edge modes of the quantum Hall effect \cite{moo}. 

It is well known that adding marginal terms to an action can drive the system to a new fixed point, describing an entirely different physics. In the present case there are indeed three additional marginal terms that can be added to the model (\ref{four}): $b_{\mu \nu} b_{\mu \nu} $, $b_{\mu \nu} F_{\mu \nu}$ and $b_{\mu \nu} \epsilon _{\mu \nu \alpha \beta } b_{\alpha \beta} $. All these terms, taken one by one, break explicitly the tensor gauge invariance (\ref{three}) and introduce thus new, unwanted degrees of freedom. There is, however one particular combination of these three terms that can be added to the effective action and that preserves both gauge invariances, albeit (\ref{three}) is realized in a different, more subtle way:
\begin{eqnarray}
S &&= {i\over 32 \theta} \int d^4x \ \left( kb_{\mu \nu}+{\theta \over \pi }F_{\mu \nu} \right) 
\epsilon_{\mu \nu \alpha \beta} \left( kb_{\alpha \beta} +{\theta \over \pi } F_{\alpha \beta} \right) 
\nonumber \\
&&+  {\pi^2 \over  4e^2\theta^2 } \int d^4x \ \left( kb_{\mu \nu} + {\theta \over \pi }F_{\mu \nu}\right) \left( kb_{\mu \nu} + {\theta \over \pi }F_{\mu \nu} \right) ,
\label{seven}
\end{eqnarray}

Something very interesting happens when the additional marginal terms in the effective action combine with the original ones to give (\ref{seven}). Suppose, moreover, that the charge current is identified with the boundaries of the now open world-surfaces of the vortices as $J_{\mu } = 2 \partial_{\nu} \Phi_{\mu \nu}$, so that the interaction can be combined into
\begin{equation}
S_{\rm int}= \int d^4 x \left(k b_{\mu \nu} + {\theta \over \pi }F_{\mu \nu}\right) \Phi_{\mu \nu} \ ,
\label{eight}
\end{equation}
describing particles of charge $\theta /\pi$ at the end of magnetic vortex lines. 
Then, only the combination $\left( kb_{\mu \nu}+{\theta \over \pi }F_{\mu \nu} \right)$ appears in the action and the tensor gauge invariance (\ref{three}) is preserved if it is combined with a corresponding shift  $A_{\mu } \to A_{\mu } - {\pi k\over \theta} \eta_{\mu }$. This combined transformation can be exploited to entirely absorb $F_{\mu \nu}$ into $b_{\mu \nu}$, giving the effective action
\begin{eqnarray}
S &&= {i \over 32 \theta} \int d^4x \ b_{\mu \nu} \epsilon_{\mu \nu \alpha \beta} b_{\alpha \beta} 
+  \int d^4 x {\pi^2 \over  4 e^2\theta^2 } b_{\mu \nu} b_{\mu \nu}  
\nonumber \\
&&+ {i\over 16\pi} \int d^4x \ b_{\mu \nu} \epsilon_{\mu \nu \alpha \beta} F_{\alpha \beta} + i \int d^4x  \ b_{\mu \nu} \Phi_{\mu \nu} ,
\label{nine}
\end{eqnarray}
where we have specialised to the case $k=1$. 

In this gauge-fixed form, the original gauge symmetry (\ref{three}) appears as broken. This is nothing else than a Higgs mechanism of the second kind for the tensor gauge symmetry ($\ref{three}$), in which a scalar longitudinal polarization "eats up" two transverse polarizations to become a massive vector \cite{que}. This mechanism can thus be viewed as the dual of the standard Higgs mechanism. The realisation of this tensor Higgs mechanism goes hand in hand with the soldering of charges and vortices described by (\ref{eight}) into a unique string-like excitation with open world-sheets, describing magnetic fluxes with charged dyons at their ends. 

In order to establish the detailed character of the resulting soldered excitation let us compute its induced action by using the explicit form (\ref{one}) for $\Phi_{\mu \nu}$. The relevant and marginal terms in this action are
\begin{eqnarray}
&&S_{QP} = {\Lambda^2 \over 4\pi} K_0 \left( {m\over \Lambda_0} \right) \int_S d^2\sigma \sqrt{g} 
+ {\Lambda^2\over 16\pi m^2} \int _S d^2 \sigma \sqrt{g} R 
\nonumber \\
&& - {\Lambda^2\over 16\pi m^2} \int_S d^2 \sigma \sqrt{g} g^{ab} \partial_a t_{\mu \nu} \partial_b t_{\mu \nu}
\nonumber \\
&&-i{\theta \over \pi} {\pi \over {1+{4\pi \over e^2}}} \nu + {e^2 m\over 8\pi^2} f \left( {m\over \Lambda_0} \right)
\int_{\it \partial S} d\tau \sqrt{ {dx_{\mu}\over d\tau} {dx_{\mu}\over d\tau}} \ ,
\label{ten}
\end{eqnarray}
where $m = e\Lambda/4\pi$, $f(x) = \int_x^{\infty} dz K_1(z)/z$ and $K_0$ and $K_1$ are Bessel functions of imaginary argument. 
The geometric quantities in this expression are defined in terms of the induced surface metric 
\begin{equation}
g_{ab} = {\partial x_{\mu}\over \partial \sigma ^a} {\partial x_{\mu }\over \partial \sigma ^b} \ ,
\label{eleven}
\end{equation}
as $g = {\rm det}(g_{ab}) = X_{\mu \nu} X_{\mu \nu}/2$ and $t_{\mu \nu} =  X_{\mu \nu}/ \sqrt{g}$. The quantity $R$ is the scalar (intrinsic) curvature of the world-surface while 
\begin{equation}
\nu = {1\over 4\pi} \int d^2\sigma \ \sqrt{g} \epsilon_{\mu \nu \alpha \beta} g^{ab} \partial_a t_{\mu \nu}
\partial_b t_{\alpha \beta} \ ,
\label{twelve}
\end{equation}
represents the (signed) self-intersection number of the world-surface. 

At this point we must remove the ultraviolet cutoff $\Lambda_0$ to obtain the quantum numbers of the soldered point excitation. To this end we need the large $x$ asymptotic behaviour of the Bessel functions appearing in the induced action,
\begin{eqnarray}
K_0(x) &&\simeq {{\rm exp}(-x) \over \sqrt{x}} \ , \qquad x \to \infty \ ,
\nonumber \\
f(x) &&\simeq {{\rm exp}(-x)\over \sqrt{x^3}} \ , \qquad x \to \infty \ . 
\label{thirteen}
\end{eqnarray}
Since $m/\Lambda_0 = {\rm const.} e$  and $K_0$ diverges for $x\to 0$, the only way to avoid an infinity when removing the ultraviolet cutoff to infinity is to let $e\to \infty$, i.e. the system is driven to the strong coupling limit when removing the ultraviolet cutoff. 
As a first consequence we see immediately that both intrinsic and extrinsic curvature terms of the string vanish in this limit. Further, since at large $e$ we have $(e^2m/8\pi^2) f(m/\Lambda_0) = (e^2/8\pi^2 \Lambda) K_0(m/\Lambda_0)$ the boundary term also vanishes if the string tension, the coefficient of the first term in (\ref{twelve}) approaches a finite limit.  When we remove the ultraviolet cutoff  we are thus left with the renormalised action
\begin{equation}
S_R= T \int_S d^2\sigma \sqrt{g} -i{\theta \over \pi} \pi \nu \ .
\label{fourteen}
\end{equation}
For $\theta/\pi = 1$ this is the action of the spinning string \cite{pol2}. As a consequence of the tensor Higgs mechanism (or dual Higgs mechanism) charges are confined by a linear potential with string tension $T$. 
It has been also shown \cite{paw} that at $\theta /\pi = 1$ the topological term representing the signed self-intersection number of the string world-sheet is just another representation of the spin factor \cite{pol1} of point particles of spin 1/2 at the ends of the string. Not only charges are confined but the constituents of the composite "mesons" acquire spin 1/2. We have thus shown how in this simple model of spontaneous tensor symmetry breaking charges and vortices are soldered in string-like composites, thereby acquiring spin.


\begin{references}
\bibitem{ave}For a review see: {\it Strongly Correlated Systems}, A. Avella and F. Mancini eds., Springer Verlag, Berlin (2012).
\bibitem{hal}F. D. M. Haldane, {\it J. Phys. C: Solid State Phys.} {\bf 14} (1981) 2585.
\bibitem{weng}Z.-Y. Weng, D. N. Sheng, Y.-C. Chen and C. S. Ting, {\it Phys. Rev} {\bf B62} (1997) 3894.
\bibitem{ye} P. Ye, C. S. Tian, X.-L Qi and Z.-Y Weng, {\it Phys. Rev. Lett.} {\bf 106} (2011) 147002. 
\bibitem{wensach}For a review see: P. A. Lee, N. Nagaosa and X.-G. Wen, {\it Rev. Mod. Phys.} {\it 78} (2006) 17; S. Sachdev, {\it Rev. Mod. Phys.} {\bf 75} (2003) 913. 
\bibitem{sen1}T. Senthil, A. Vishwanath, L. Balents, S. Sachdev and M. Fisher, {\it Science} {\it 303} (2004) 1490. 
\bibitem{dst1}M. C. Diamantini, P. Sodano and C. A. Trugenberger,  {\it  Nucl. Phys.} {\bf B474} (1996) 641.
\bibitem{birmi}For a review see: D. Birmingham, M. Blau, M. Rakowski and G. Thompson, {\it Phsy. Rep.} {\bf 209} (1991) 129. 
\bibitem{sent}T. Senthil and M.P.A. Fisher,  {\it Phys. Rev} {\bf B62} (2000) 7850.
\bibitem{que} F. Quevedo and C. A. Trugenberger, {\it Nucl.Phys.} {\bf B501} 143 (1997). 
\bibitem{paw}J. Pawelczyk, {\it Phys. Lett.} {\bf B311} (1993) 98.
\bibitem{pol1}A. Polyakov, in {\it Fields, Strings and Critical Phenomena}, Proc. Les Houches Summer School 1988, E. BrŽzin and J. Zinn-Justin eds., North-Holland, Amsterdam (1990). 
\bibitem{geo} For a review see: H. Georgi, {\it Lie Algebras in Particle Physics}, Addison-Wesley, Redwood City (1982).
\bibitem{wil}F. Wilczek, {\it Phys. Rev. Lett.} {\bf 58} (1987) 1799. 
\bibitem{jackiw}R. Jackiw and S. Templeton, {\it Phys. Rev.} {\bf D23} (1981) 2291; 
S. Deser, R. Jackiw and S. Templeton, {\it Phys. Rev. Lett.} {\bf 48} (1982) 975;  {\it Ann. Phys.} (N.Y.) {\bf 140} (1982) 372.
\bibitem{bow}T. J. Allen, M. Bowick and A. Lahiri, {\it Mod. Phys. Lett.} {\bf A6} (1991) 559; M. Bergeron, G. Semenoff and R. J. Szabo, {\it Nucl. Phys.} {\bf B437} (1995) 695. 
\bibitem{topins}For a review see: M.Z. Hasan and C. L. Kane, {\it Rev. Mod. Phys} {\bf 82} (2010) 3045; M. Z. Hasan and J. E. Moore, {\it Ann. Rev. Cond. Matt. Phys.} {\bf 2} (2010) 55 AOP. 
\bibitem{moo} G. Y. Cho and J. E. Moore, {\it Ann. Phys.} {\bf 326} (2011) 1515.
\bibitem{pol2}A. Polyakov, {\it Gauge Fields and Strings},  Harwood Academic Publishers, Chur (1087).

\end{references}
\end{document}